# AC-Biased Shift Registers as Fabrication Process Benchmark Circuits and Flux Trapping Diagnostic Tool

Vasili K. Semenov, Yuri A. Polyakov, and Sergey K. Tolpygo

*Abstract*—We develop an ac-biased shift register introduced in our previous work (V.K. Semenov *et al.*, *IEEE Trans. Appl. Supercond.*, vol. 25, no. 3, 1301507, June 2015) into a benchmark circuit for evaluation of superconductor electronics fabrication technology. The developed testing technique allows for extracting margins of all individual cells in the shift register, which in turn makes it possible to estimate statistical distribution of Josephson junctions in the circuit. We applied this approach to successfully test registers having 8, 16, 36, and 202 thousand cells and, respectively, about 33, 65, 144, and 809 thousand Josephson junctions (JJs). The circuits were fabricated at MIT Lincoln Laboratory, using a fully planarized process, 0.4 µm inductor linewidth and $1.33 \cdot 10^6$ cm$^{-2}$ junction density. They are presently the largest operational superconducting SFQ circuits ever made. The developed technique distinguishes between "hard" defects (fabrication-related) and "soft" defects (measurement-related) and locates them in the circuit. The "soft" defects are specific to superconducting circuits and caused by magnetic flux trapping either inside the active cells or in the dedicated flux-trapping moats near the cells. The number and distribution of "soft" defects depend on the ambient magnetic field and vary with thermal cycling even if done in the same magnetic environment.

## I. INTRODUCTION

SUPERCONDUCTOR digital devices are clear winners of several types of electronics benchmarks - some devices demonstrated the highest, up to ~750 GHz, clock rate [1], whereas other devices showed the lowest, below $10^{-11}$ pJ, energy dissipation per logic operation [2]. These and many other impressive results were achieved using the device fabrication tools antiquated from the standpoint of modern microelectronics. The evident mismatch between the outdated tools and record-setting results stimulated many speculations about the potential of superconductor digital circuits made using modern microelectronics foundries. The IARPA C3 Program [3] gives us a chance to correlate the fabrication progress [4], [5] and complexity of superconducting circuits [6]. Prototypes of superconducting microprocessors and their components presented at ASC 2016 [7], [8] [9], [10] show new opportunities for rapid development of digital superconductor electronics.

## II. STRUCTURE OF THE TEST CIRCUIT

Benchmark test circuits are highly desirable as a tool to evaluate the maturity of fabrication processes. Benchmark circuits can play an important role in distinguishing between numerous potential fail factors. Because of various reasons, only a few types of benchmark circuits and process diagnostic tests have been published and discussed. Shift registers dominate the pool of benchmark circuits due to a good balance between their universality and simplicity [11],[12]. Random access memories are second due to possibility of investigating partly operational circuits with defective cells and extracting their exact locations [13].

In our previous work, we suggested a simple ac-biased shift register and explained how it can be used for generic technology benchmarking [14]. It is kind of a "scan chain" wildly used in CMOS technology. On the one hand, our circuit shares drawbacks of all shift registers – it is a periodic structure and does not have "random" logic. On the other hand, it offers a unique possibility of evaluating critical currents of Josephson junctions in every register cell.

A unit cell of the register developed in [14] is shown in Fig. 1. Each cell contains a chain of four inductances (L1 - L4), each connected to the ground plane by a Josephson junction. All four junctions (J1 – J4) have identical target parameters. The cell also has a "handle" - inductance L6 inductively coupled to the common ac clock line. Since [14], we implemented a number of cell improvements driven by a progress with understanding technological limitations. Besides, we have made two major layout upgrades of the original layout [14] shown in Fig. 1(c). Fig. 1(d) illustrates our progress with cell miniaturization due to incremental upgrades of the MIT LL process from the SFQ4ee to the SFQ5ee node. In particular, we used a smaller number of larger 'dummy' Josephson junctions and more aggressive inductor linewidth and via design rules. Fig. 1(e) shows the layout modified for a prospective technology with self-shunted Josephson junctions.

This work was supported in part by the Office of the Director of National Intelligence (ODNI), Intelligence Advanced Research Projects Activity (IARPA) via Air Force Contract FA872105C0002. The views and conclusions contained herein are those of the authors and should not be interpreted as necessarily representing the official policies or endorsements, either expressed or implied, of the ODNI, IARPA, or the U.S. Government. The U.S. Government is authorized to reproduce and distribute reprints for Governmental purposes notwithstanding any copyright annotation thereon.

V.K. Semenov and Y.A. Polyakov are with the Department of Physics and Astronomy, Stony Brook University, Stony Brook, NY 11794, USA (e-mail: Vasili.Semenov@StonyBrook.edu; Yuri.Polyakov@StonyBrook.edu).

S.K. Tolpygo is with the Lincoln Laboratory, Massachusetts Institute of Technology, Lexington, MA 02420 USA (e-mail: sergey.tolpygo@ll.mit.edu).

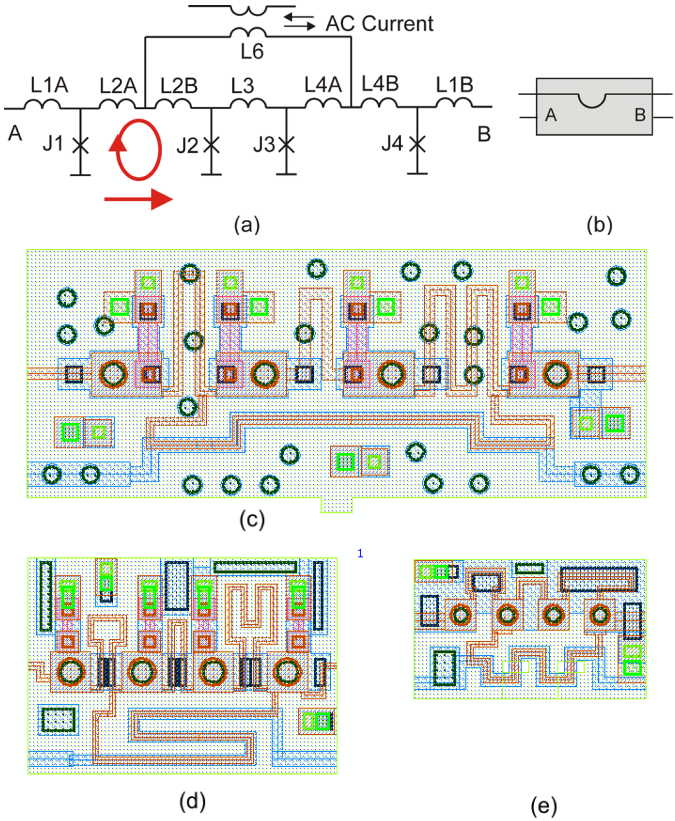

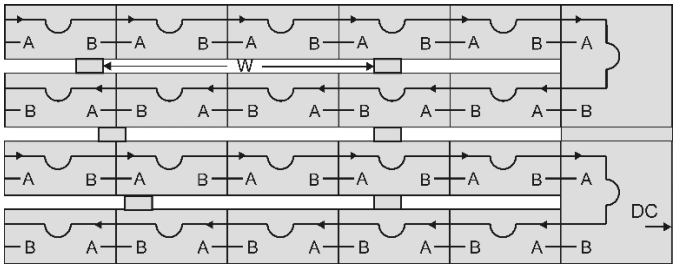

Fig. 1. Schematics (a), notation (b), and layouts (c)-(e) of shift register (flux shuttle) cells. Cell dimensions (including 1-μm-wide moat) are 17 μm x 40 μm = 680 μm² for layout (c), 15 μm x 20 μm = 300 μm² for layout (d) and 10 μm x 15 μm = 150 μm² for layout (e). Inductance linewidth in (d),(e) was 0.4 μm.

Fig. 2. Meander structure of the ac-biased shift register. Two types of "u-turn" cells connect right ends of the horizontal strings of unit cells shown in Fig. 1.

Regular cells are connected into interleaved strings with opposite data propagation directions (Fig. 2). The ends of neighboring strings are connected by "u-turn" cells (the upper right corner in Fig. 2), which are functionally and schematically identical to the regular cells shown in Fig. 1. In [14] we demonstrated a possibility to tap data from internal points of the register using dc-biased RSFQ-type "u-turn" inserts. In this work, we use a more efficient way to tap data by using a new ac-biased u-turn cell shown in Fig. 3. We designed the cell in two steps. Firstly, we connected junction J3 with the first junction (J5) of the usual (dc-biased) JTL line via a relatively large (2 PSCAN units or 5.28 pH) inductance. Then we used PSCAN to align margins of the u-turn cell with margins of the unit cells by adjusting the cell inductances.

We placed the strings of cells (Fig. 2) at 1-μm spacing. The empty spaces between them make very long moats. We cut them into shorter moats by connecting the sky and ground planes of the adjacent rows of cells by short bridges. The distance $W$ between the bridges does not affect the circuit operation but could affect the moat efficiency. Different values of $W$ could be used within a single register for a comparative study of moat efficiencies.

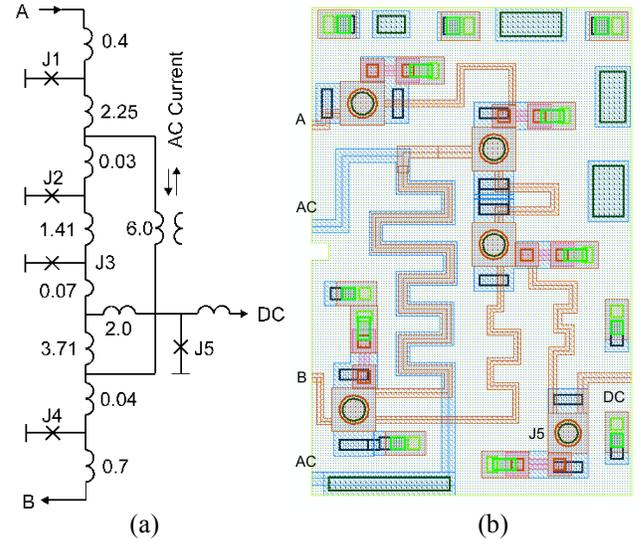

Fig. 3. Schematics (a) and layout (b) of the optimized tap for reading data from the ac-biased cell to a dc-biased SFQ-DC monitor. Inductances are shown in PSCAN units $L_{PSCAN}$ = 2.64 pH; critical currents of J1 – J4 junctions are 0.125 mA, critical current of J5 is about 0.1 mA.

### III. MEASUREMENT TECHNIQUE

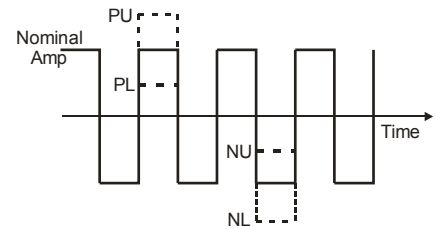

Fig. 4. Rectangular clock pulses. Positive or/and negative amplitude of a set of clock pulses can be changed with respect to the nominal in order to probe the margins of the corresponding cells.

Before extracting margins of $N$ individual cells, we should check that the $N$-bit register is operational and find its global margins – a range of positive and negative amplitudes of rectangle clock pulses shifting the input data pattern from the register input to its output after $N$ clock periods. These correct or "nominal" amplitudes will be extensively used to write in and read out the special test patterns. Our technique essentially uses the fact that margins of empty cells (that store logic "zeros") are much wider than margins of cells with magnetic flux quanta that store logic "ones". The technique is easier to explain using a simplest pattern containing only a single logic "1" written into a cell $k$ under investigation. The main idea of the method is to shift this "1" to the next position using one clock pulse with *intentionally modified* amplitude, either positive or negative, as shown in Fig. 4. To check the result, we read out the register content using the nominal amplitudes and compare it with our expectations. The circuit operates correctly, if the "1" in the cell $k$ is shifted into the next, $k+1$ cell. No shift or shifts to more than one position are considered as errors. To find the operation margins of the cell $k$ we should repeat the described procedure using different modified amplitudes. Then the whole test cycle should be repeated for all $N$ cells.

The described ideal pattern is too slow to be practical for long registers. The procedure can be accelerated by using patterns with sparse "1"s. For example, the extraction of margins can start using a pattern with logic "1"s separated by two "0"s: 1 0 0 1 0 0 1 0 … This pattern deals with the first, fourth, seventh, and so on cells. In particular, we used it to identify and extract margins of the weakest cell among the mentioned. Then, the identified cell is eliminated from further measurements by replacing "1" corresponding to the cell by "0". For example, if the weakest cell was in position four, the new pattern becomes 1 0 0 0 0 0 1 0 … The new pattern with the reduced number of "1"s is then used to identify the cell with the second weakest margins, and so on. The described procedure continues until all margins are extracted and the pattern contains only "0"s. Then, the whole sequence is repeated using a "shifted" sparse pattern: 0 1 0 0 1 0 0 1 … that deals with cells in the second, fifth, eighth, and so on positions. Finally, the procedure is repeated with pattern 0 0 1 0 0 1 0 0 1 …

Using the described procedure, we measured margins of cells for both polarities of ac clock. This information can be also interpreted in terms of deviations of critical currents in shift register Josephson junctions. Indeed, effect of any cell distortion can be extracted by straightforward numerical simulations of the cell margins with the nominal and distorted values of the selected parameter. Table I shows the results when the distorted parameters are critical currents of four Josephson junctions. The dimensionless numbers in Table I are the rates of margin changes with respect to changes of the critical currents. According to Table I, variations of critical current of junction J2 have the major impact on the lower positive margin. Numerically it is about 4.4 μA change of the margin per 1 μA change of the critical current. Similarly, junction J4 has the largest impact on the upper negative margin. Impacts of simultaneous deviations of several critical currents can be also calculated using linear coefficients in Table I. For example, if we assume equal and statistically independent deviations of all four junctions then the cumulative impact of deviations could be described by two coefficients shown in the last column of Table I. These dimensionless coefficients connect rms deviations of critical currents with rms deviations of clock margins. The coefficients are close to each other. With a reasonable accuracy, it is possible to state that, if the measured spread of clock margins $rms_{AMP}$ is caused by the spread of critical currents $rms_{IC}$, then the spread of critical currents is about 5.5 times lower than the spread of the clock amplitudes.

$$rms_{ABM} \approx 5.5 \cdot rms_{IC} . \qquad (1)$$

## IV. MEASUREMENT RESULTS

### A. Smaller shift register with 16k cells

A number of shift registers have been laid out, fabricated at MIT Lincoln Laboratory using the SFQ4ee and SFQ5ee process nodes with 100-μA/μm² Nb/AlO$_x$/Al/Nb Josephson junctions [4],[5], and investigated. As mentioned earlier [14], it is possible to extract the lower and upper operation margins for ac clock current of two memory loops - between junctions J1 and J2 and between junctions J3 and J4 (see Fig. 1) - for every cell in the register, thus characterizing the uniformity of the fabricated cells. Positive clock amplitudes shift logic "1" from the first memory loop to the second. Negative clock amplitudes shift logic "1" from the second memory loop, between J3 and J4, to the first memory loop of the next cell. This tremendous amount of information about individual cells can be highly useful for characterizing and optimizing the fabrication processes.

Fig. 5 shows a few types of margins for a 16,384-bit register. Fig. 5(a) shows margins for the first memory loop, between J1 and J2. To simplify the plot, the upper individual margins were replaced by an open rectangle corresponding to the worst value of the upper individual margins. Lower individual margins of edge cells (rows 1 and 128) are visibly lower than the margins of all other cells. The old-fashion dc-biased u-turn inserts used in this particular register cause this. It was the reason to redesign them as shown in Fig. 3. Margins for the internal cells are practically flat. Fig. 5(b) shows a blowup of the flat area. The next two plots are the general view of margins, Fig. 5(c), and the blowup, Fig. 5(d), for the second memory loops, showing similar to the first loop statistical properties.

To sort out possible origins of the spread of individual margins, we repeated the full cycle of measurements with and without thermal cycles. The thin blue curve in Fig. 6(a) shows sorted lower positive margins averaged over 10 thermal cycles. The thicker red curve is the best fit of the data by an error function defined by the mean value (margin center) and standard deviation σ. The fitted value σ = 14.8 μA is equivalent to 1-σ spread of the critical currents of junctions in the circuit of 2.7 μA.

The latter value characterizes a cumulative impact of the fabrication spread and, to a lesser degree, impacts of flux trapping and random measurement errors. This standard deviation is quite close to the typical fabrication spread of junctions with 0.125 mA critical currents [5]. It does not leave much room for random measurement errors and flux trapping effects. Nevertheless, we estimated also these factors.

The measurement errors can be extracted by analyzing the difference between one set of measurements and the averaged data, as shown in Fig. 6(b). The error of measurements (1-σ deviation) of the amplitude is only 2.6 μA. The equivalent spread of critical currents according to (1) is only about 0.47 μA. This very low value, comparable to thermally induced uncertainty of critical currents, can serve as a conservative upper limit to the accuracy of our measurements.

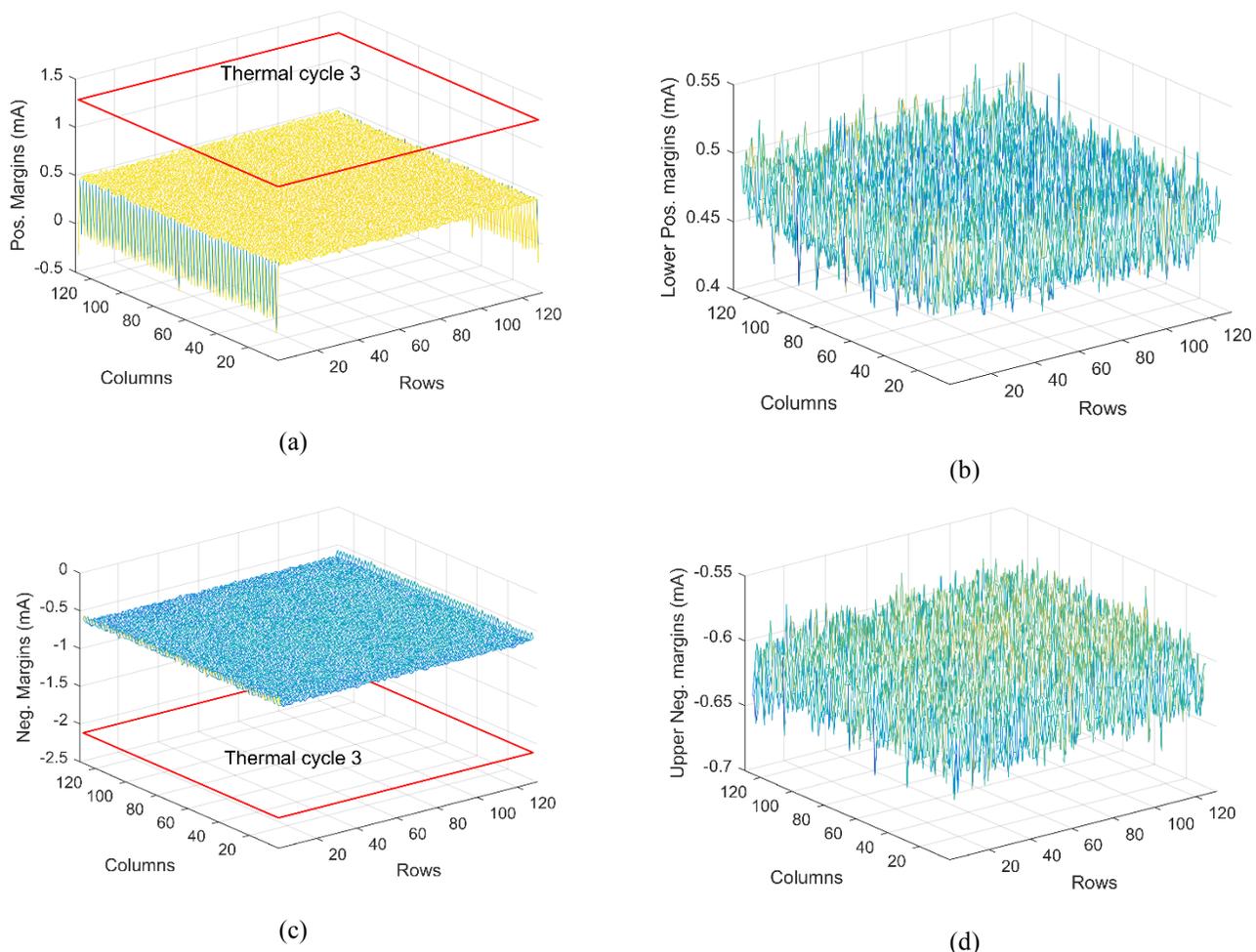

Fig. 5. Common and individual cells margins of the 16k register. The upper and lower rows of plots accordingly show margins for the first memory loop with junctions J1 and J2 (a), (b), and for the second memory loop with junctions J3, J4 (c), (d). Left and right columns show accordingly margins with coarse (a), (c) and fine (b), (d) resolutions. Open red rectangles show accordingly common (worst) upper positive (a) and lower negative (c) margins.

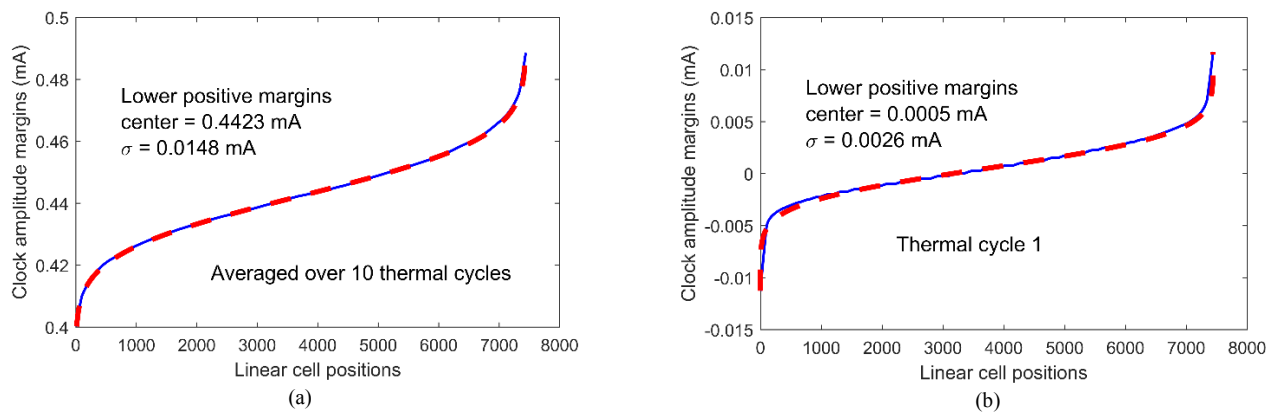

Fig. 6. Statistical properties of the lower half of the 16k register. Sorted measured margins (blue solid curves) and their best fits by error functions (wider red dash curves) for lower positive margins of individual cells averaged over 10 thermal cycles (a); and deviations of the margins after the first thermal cycle from their values averaged over 10 cycles (b).

Despite lack of outliers, we can suggest that impact of flux trapping is associated with the largest deviations of margins shown in Fig. 6(b) from zero. Fig. 7 shows locations of cells with the largest deviations of lower positive margins measured after each of 10 thermal cycles. There are a few coinciding locations but the majority of them look quite random.

### B. A larger shift register with 202,280 cells

Comprehensive measurements of robust 8k, 16k and 36k shift registers were convenient for investigations of small trapping effects by means of comparative study of circuits exposed to thermal cycles. Measurements of longer resisters are more difficult and time consuming. We discuss here only the measurement carried out within one thermal cycling. The register (Fig. 8) occupies 8 mm x 8 mm payload area of a 1 cm x 1 cm chip. Some small payload area was reserved for two 16-bit registers, two Josephson junctions with 0.25 mA critical currents that served as thermometers and two 40-Ω resistors that served as local heaters. The main shift register consists of 202,280 cells shown in Figs. 1(a,d), and 3. The register contains one data input, one data output, and 17 intermediate data taps that allow to diagnose partly operational circuits and simplify the debugging and optimization of the measurement procedure. The measurements were complicated by much longer time required for working with longer data patterns and stronger requirements for reliability of the testing procedures. In particular, the achieved 0.01% error rate for a single bit measurement event was insufficient for our purpose. To resolve the problem we repeated all measurements ten times and then analysed the redundant data to exclude incorrect measurement results. It was not so difficult because one or two incorrect values were very different from groups of 9 or 8 very close correct values. Only one of six investigated chips was fully operational, while the other five were alive but only partly operational.

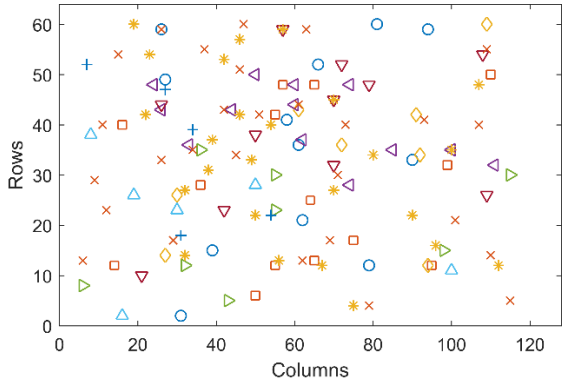

Fig. 7. Positions of cells with largest deviations of lower positive clock margins. Colors and marker types identify thermal cycle numbers. These positions characterize the distribution of magnetic flux trapped in the circuit moats after each thermal cycle.

TABLE I
RATES OF CHANGE OF REGISTER CELL MARGINS WITH RESPECT TO CHANGES OF CRITICAL CURRENTS OF JOSEPHSON JUNCTIONS

|  | Ic1 | Ic2 | Ic3 | Ic4 | Cum. impact |
|---|---|---|---|---|---|
| $d(posMarg)/dI_{ci}$ | -1.93 | 4.40 | 2.46 | 0 | 5.40 |
| $d(negMarg)/dI_{ci}$ | 0 | -0.97 | -1.14 | 5.37 | 5.58 |

As the first measurement step, we applied to the resister a short (with six logic "1"s and six logic "0"s) "active" pseudo random pattern followed by the sufficient number of "0"s to match the pattern and register lengths. Several taps were used to monitor the propagation of the pattern through the register. Output patterns were automatically compared with the input ones. To find the operation range, we automatically scanned parameters of interest and repeated the testing procedure at each selected set of parameters. Three plots in Fig. 9 show different grades of register operations. Blue open circles mark correct operations; red squares mark operations with correct patterns but with incorrect (shorter) delays between input and output patterns. Finally, green diamonds mark operations with correct number of logical "1"s but with distorted distances between them.

The leftmost plot shows "the perfect" operation of the shortest register section containing only 389 cells in a very wide ranges of amplitudes from about 0.5 to 2.2 mA for positive and -0.5 to -2.5 mA for negative amplitudes. The middle plot shows a noticeable operation area of a register section with 24,896 cells. The rightmost plot shows that the register is still operational but the delay between input and output data is always shorter than the expected for this tap 75,077 clock periods. Each thermal cycle changed the operational areas in the register and even the existence of the operation grades.

Shorter delays mean that a tiny fraction of the register cells operate as pieces of Josephson transmission lines (JTLs). This type of errors is typically observed when the absolute values of clock amplitudes are too high for the correct operations. For shorter registers such a behavior was not an issue because we were able to return to the normal operation by reducing the clock amplitudes. For the sections of the longer register, we could not do this because some other "bad" cells stopped to work at the reduced amplitudes.

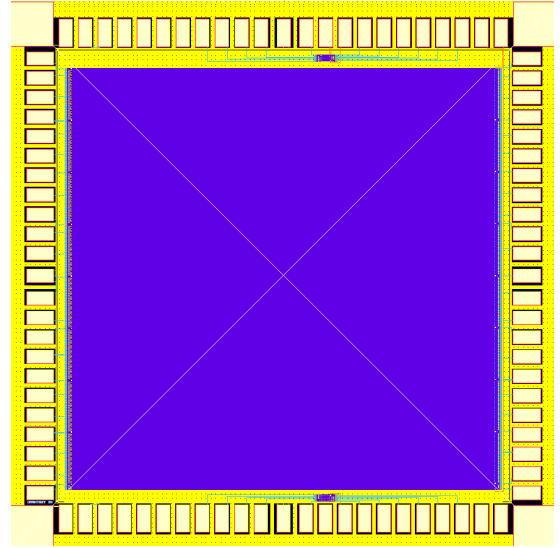

Fig. 8. Floor plan of 1 cm x 1 cm chip with one large shift register, two small shift registers, two heaters, and two Josephson junctions that were used as thermometers.

The described effect would be unacceptable for practical circuits. However, the observed shortening of the delay is somehow fruitful because it allows us to diagnose the technology. For example, we learned that the measured 202,218 clock period length of the register was by 62 clock period shorter than its designed 202,280 clock period length. In other words, 62 cells operated as JTLs and they were hidden from our testing procedure. We will refer to them as missing cells. The ratio of missing and operational cells, about 0.03% in this case, is a way to characterize the technology. The number of missing cells usually changed after each thermal cycle. It means that the missing-cell effect originated from the flux frozen in some unexpected places. Fortunately, the described measurement technique works well for registers with missing cells.

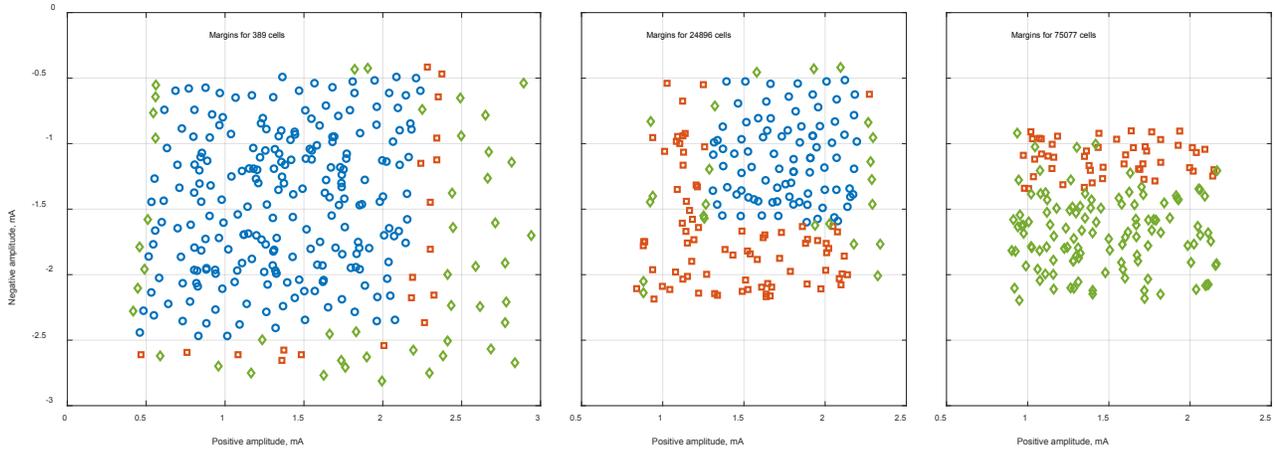

Fig. 9. Degradation of margins with the shift register length. Blue circles mark the full matching between the measured and expected patterns; red squares mark incorrectly delayed but undestorted measured pattern; green diamonds mark operation with distorted distances between logic "1"s but with the correct number of logic "1"s.

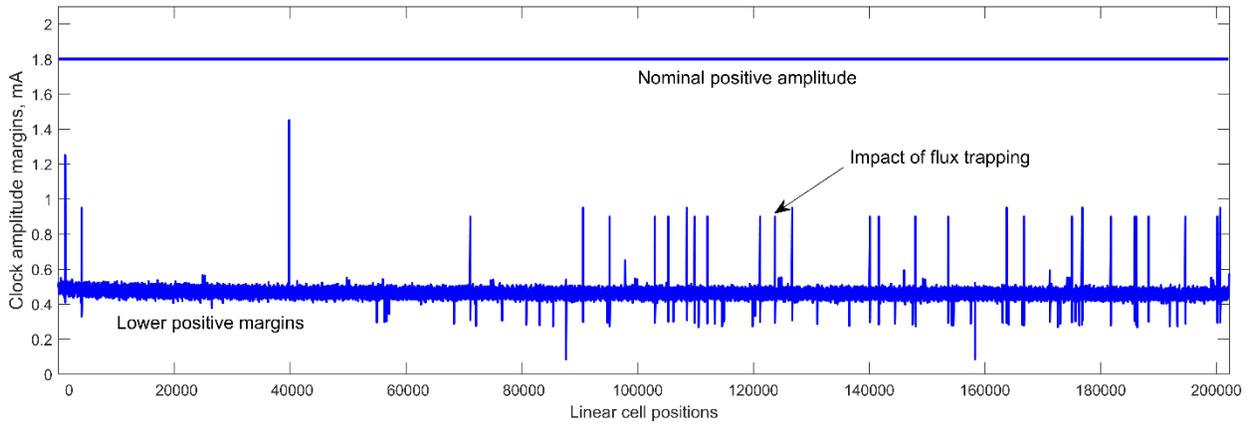

Fig. 10. Lower positive ac clock margins for accesible 202,218 cells. Horizontal line shows nominal positive amplitude of the ac clock signals used during the extraction of clock margins of the cells.

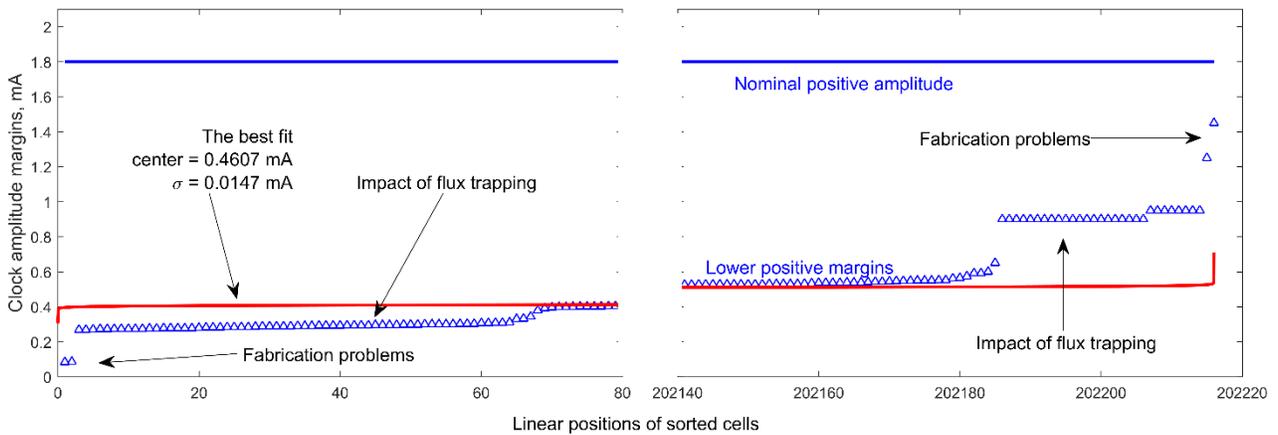

Fig. 11. The same as Fig. 10, but the cell margins were sorted in ascending order from left to right. The fit to Guassian distribution (error function) of margins is shown by the red curve, giving the mean value of 0.4607 mA and the standard deviation of 14.7 µA. This corresponds to a 1-σ spread of the critical currents of junctions of about 3 µA. Non-gaussian tails correspond to cells impacted by flux trapping in or near the cells and to four likely fabricaiton-related distortions of the cells, as shown in the Fig.

Fig. 10 shows lower positive clock margins. These data are similar to lower positive margins shown in Fig. 5(a). However, in Fig. 10 we use one-dimensional numbering showing the "clock period" distance of cells from the register input. In this way it is easy to see a number of rare but really large spikes of margins. At the first glance they look random and difficult to organize. However, sorting them in the ascending order does the trick, as shown in Fig. 11. It shows only the tails of the distribution, showing a noticeable deviation from the mean value and containing less than 0.05% data points. The 99.95% of the data in the center of the plot were omited for clarity because they match the Gaussian distribution with the accuracy better than the width of lines in the plot.

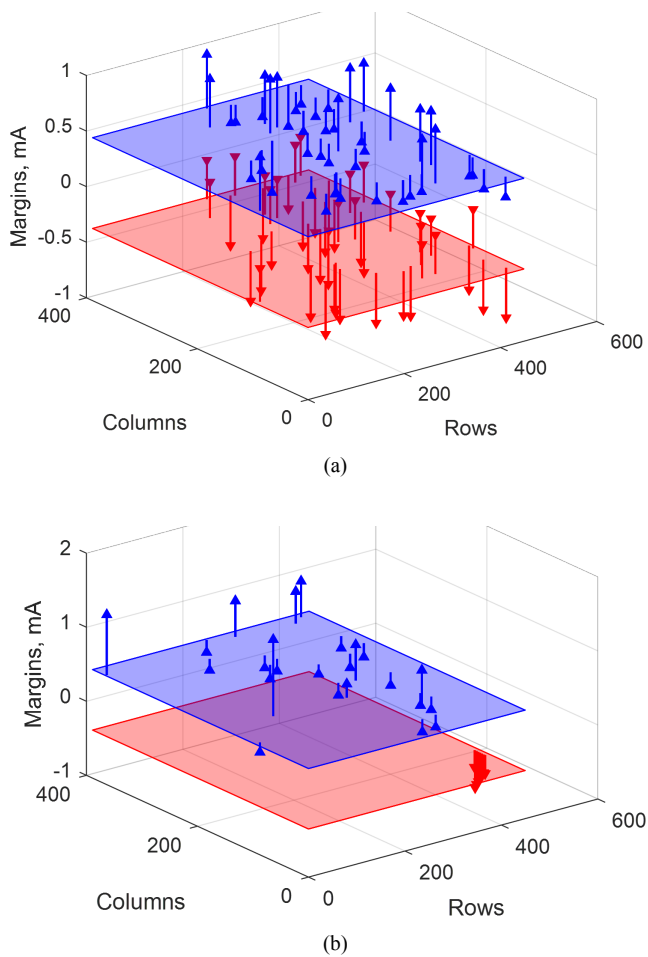

Fig. 12. Cell maps showing positions of cells with "bad" margins. Most of the instances of bad positive and bad negative margins occur either in the same or adjacent cells (a). Much fewer cells, located remotely, show no correlation between bad positive and bad negative margins (b).

The tails contain spikes that can be sorted into a few groups. The first group contains only two lowest and two highest margin data points that could be related to the cells with the fabrication-related distortions of parameters. The other two, more populated, groups are probably caused by fluxes frozen in unexpected locations, but we cannot prove it now. Note that the number of cells affected by the flux trapping and shown in the left side of Fig. 11 exactly coincides with the number of "missing" cells. We believe that both observed effects are correlated. In other words, each "missed" cell disturbs margin of its neighbor cell. Extracted ~15 µA 1σ-spread of clock amplitudes and corresponding ~3 µA spread of the critical currents of more than 809k junctions in the register are identical to these parameters in the shorter, 16k-bit register fabricated six months earlier.

Fig. 12(a) shows that most spikes in positive margins in Fig. 10 strongly correlate with spikes in negative margins in the same or adjacent cells. Fig. 12(b) shows complementary data showing positions of cells with uncorrelated positive and negative spikes in margins. Comparison of these Figs. shows that collective imperfections that involve either both memory loops of one cell or memory loops of neighboring cells are more common than local defects affecting single circuit components. The data shown in Fig. 12 were taken from an intermediate tap #15 in the register, probing about 175 thousand cells. This difference should be taken into account when comparing the data in Fig. 12 and Fig. 10.

## V. CONCLUSION

The-state-of-the-art in superconductor digital technology is close to the psychologically important million-junction level of integration. We have shown in this work that circuits with such level of integration can be designed, fabricated, and successfully tested. The main advantage of our ac-biased shift register is a possibility to extract properties of individual cells. In this respect, our circuits are similar to random access memories. However, in contrast to RAMs, our circuits are simpler for design and take less man-power to layout. The ability to access individual cells is especially important for superconductor circuits that could be affected by parasitic flux trapping, the phenomenon nonexistent in semiconductor electronic circuits.

We have compared the properties of circuits with different integration levels, from about 33k junctions to more than 809k junctions per circuit. We have found that 1-σ spread of critical currents (~3 µA) does not depend on the level of integration and hence can be measured using small circuits. In contrast, we have detected some fabrication defects with very low, ~ $5·10^{-6}$ defect per Josephson junction, probabilities which could have been detected only using circuits with the maximum possible level of integration.

We have definitely observed a significant difference in the probabilities of high impact flux-trapping events in the relatively small registers on 5-mm chips and the largest registers on 1-cm chips. The rate of high impact flux-trapping in the larger circuits is about one event per 2,000 cells. However, we did not find even a single similar effect after 10 thermal cycles of the circuit with over 16,000 cells. We cannot suggest a simple explanation of this difference in flux trapping.

It is possible however that the difference is a result of a high sensitivity of flux trapping to the fabrication-dependent flux "freezing temperature" of superconducting films, as noted in [15]. Besides, the known theoretical flux trapping investigations dealt only with simple single-layer film structures. The real digital circuits contain from 6 to 9 layers of metallization. Properties of such circuits are extremely sensitive to differences between critical temperatures of superconducting layers and critical temperatures of vias connecting them.

We believe that the practical multilayer structures could and should be analyzed theoretically despite of these complications. The practical value of the theoretical modeling of the flux trapping may be limited because of a large number of combinations of various, often not fully known, parameters and high sensitivity to some of them. It could well happen that experimentation with flux trapping in multilayer structures will be a more practical solution to the problem. We are confident that test circuits for analyzing flux trapping effects must be a common test structure, similarly to those that measure resistive properties of normal and inductive properties of superconductor films. The circuits and techniques developed in this work are significant steps in this direction.

ACKNOWLEDGMENT

We would like to thank M. Manheimer, G. Gibson, and S.V. Rylov for the stimulating discussions; V. Bolkhovsky, T. Weir, and A. Borodin for their help; E. Dauler, L.M. Johnson, and M.A. Gouker for their interest and support of this work; C. Fourie, M. Volkmann, and M.M. Khapaev for support of their unique software packages.